Title:

# Digitally Capturing Physical Prototypes During Early-Stage Engineering Design Projects for Initial Analysis of Project Output and Progression

List of authors:

- Jorgen F. Erichsen
- Heikki Sjöman
- Martin Steinert
- Torgeir Welo

Authors' Affiliation:

Norwegian University of Science and Technology (NTNU),

Department of Mechanical and Industrial Engineering

(all four authors share the same affiliation)

Corresponding Author:

Name: Jorgen F. Erichsen

Email: jorgen.erichsen@ntnu.no

Address: Richard Birkelands Veg 2B, 7491 Trondheim, Norway

Short Title:

Digitally Capturing Physical Prototypes

Number of Manuscript Pages:   27
Number of Tables:             2
Number of Figures:            9

## Abstract

Aiming to help researchers capture output from the early stages of engineering design projects, this article presents a new research tool for digitally capturing physical prototypes. The motivation for this work is to collect observations that can aid in understanding prototyping in the early stages of engineering design projects, and this article investigates if and how digital capture of physical prototypes can be used for this purpose. Early-stage prototypes are usually rough and of low-fidelity and are thus often discarded or substantially modified through the projects. Hence, retrospective access to prototypes is a challenge when trying to gather accurate empirical data. To capture the prototypes developed through the early stages of a project, a new research tool has been developed for capturing prototypes through multi-view images, along with metadata describing by whom, why, when and where the prototypes were captured.

Over the course of 17 months, this research tool has been used to capture more than 800 physical prototypes from 76 individual users across many projects. In this article, one project is shown in detail to demonstrate how this capturing system can gather empirical data for enriching engineering design project cases that focus on prototyping for concept generation. The authors also analyse the metadata provided by the system to give understanding into prototyping patterns in the projects. Lastly, through enabling digital capture of large quantities of data, the research tool presents the foundations for training artificial intelligence-based predictors and classifiers that can be used for analysis in engineering design research.

# 1 Introduction and Background

This article addresses the challenging task of studying prototyping in engineering design projects at their early design stages, and proposes to do so by digitally capturing physical prototypes. The broad objective of this work is to enable gathering more observations from early-stage engineering design projects, thus aiding engineering design researchers in studying these projects. This article proposes a new research tool for capturing project output through digitally capturing physical prototypes and presents a solution as to how to implement such a tool in engineering design research scenarios. This example implementation is then demonstrated and discussed in detail.

Prototyping is identified as a core activity of engineering design, and has been a relevant research topic for decades (Gero & Lindemann, 2005). Moreover, Wall, Ulrich, & Flowers (1992) name prototyping as "one of the most critical activities of new product development. Yet, the 19 different definitions of 'prototype' (i.e. the artefact) summarized by L. S. Jensen, Özkil, & Mortensen (2016) indicate that there is still much to understand and research regarding both prototypes and prototyping in engineering design.

In this article, *prototyping* is defined as the activity of designing, building, testing and evaluating various concepts and ideas during the design process. Using this definition, prototyping primarily yields physical output in the form of prototypes, and tacit output in the form of knowledge, skills and insights for the design team. Therefore, prototypes (i.e. the artefacts) are considered output from prototyping (i.e. the activity) in this article. This differs from most of the 19 definitions listed by L. S. Jensen et al. (2016), who simply states that prototyping is the activity of making prototypes. In broader terms, prototyping could be categorised under design activity (Erichsen, Wulvik, Steinert, & Welo, 2019).

## 1.1 Research on Prototyping and Design Activity

How design activities are commonly researched in early-stage engineering design projects is of key importance. Unfortunately, the availability of literature on empirical studies of prototyping activity that specifically targets early-stage engineering design projects is scarce (Erichsen et al., 2019). There are other studies targeting the quantification of prototyping, though these typically target later stage of the design process. Studies that investigate prototyping often capture and analyse audio and video (Cash, Hicks, Culley, & Salustri, 2011; Jung, Sirkin, Gür, & Steinert, 2015; Sonalkar, Jablokow, Edelman, Mabogunje, & Leifer, 2017) or protocols (Ahmed & Christensen, 2009; Ball & Christensen, 2009, 2018; Christensen & Schunn, 2007; Dorst & Cross, 2001; Mabogunje, Eris, Sonalkar, Jung, & Leifer, 2009) from design sessions. Access to industrial projects and participants is limited, and a large portion of such studies only include a handful of participants (Ball & Christensen, 2009, 2018; Cash et al., 2011; Mabogunje et al., 2009). To circumvent this accessibility issue, many studies employ student participants— either in experiments (Ariff, Badke-Schaub, Eris, & Suib, 2012; Cash & Maier, 2016; Dong, 2005; Dong, Hill, & Agogino, 2004; Eris, Martelaro, & Badke-Schaub, 2014; Gonçalves, Cardoso, & Badke-Schaub, 2012; Jung, Martelaro, & Hinds, 2015; Larsson, Törlind, Mabogunje, & Milne, 2002; Sonalkar et al., 2017) or through investigating university course deliverables (V. Viswanathan, Atilola, Esposito, & Linsey, 2014) and logbooks (McAlpine, Cash, & Hicks, 2017).

The research articles using industry cases are often retrospective, using the outcome from the prototyping (i.e. the prototypes and project outcome) as a basis, e.g. Shah, Smith, & Vargas-Hernandez (2003) and Wall et al. (1992). Moreover, research articles that report on semi-controlled experiments often conduct performance assessments on given design criteria (V. Viswanathan et al., 2014) or perform exhaustive manual coding of transcripts of short design sessions to generalise on designer behaviour (Ball and Christensen 2009). In semi-controlled experiments, there tends to be a mix of analysing temporal data and retrospective data, as the sessions are often shorter.

In retrospective engineering design case studies, the researcher would have to perform protocol studies (Chakrabarti, Morgenstern, & Knaab, 2004), investigate logbooks (McAlpine et al., 2017; Yang, 2009) or do interviews with the project participants, and would have to establish this timeline based on the recollection of the participants and fidelity the of protocols. Minor, yet still important, modifications might be overlooked when using such methods, as the prototypes often evolve over time. Moreover, prototypes are often forgotten or lost because they are not formalised before being reused, potentially losing out on valuable information.

## 1.2 Capturing output from activity (as opposed to capturing the activity itself)

In the early, pre-requirement stages of design, output from prototyping is often low-fidelity, inexpensive prototypes (Auflem, Erichsen, & Steinert, 2019; Lauff, Kotys-Schwartz, & Rentschler, 2018). These prototypes are interesting because—as stated by Auflem et al. (2019)—they often contribute to insights for the design team. However, a common problem for engineering design researchers is access to such prototypes, especially since research on prototypes is mostly done retrospectively, and requires great effort (Erichsen et al., 2019). Studying later-stage design activities is possible through e.g. Computer Aided Design (CAD) applications (Ishino & Jin, 2002), and efforts improving such interactions is demonstrated by e.g. (Raffaeli, Mengoni, & Germani, 2013).

Wall et al. (1992) evaluate output from prototyping technologies in an industrial context. Notably, the data gathered in this study are drawn from surveys and their estimation of prototype properties, not from investigating the prototypes themselves. By investigating this study, it is hard to judge the selection criteria of the plus 100 selected prototypes and whether they all belong in the same product portfolio.

A practical challenge with studying prototypes is that storing physical prototypes from early-stage development projects is often impractical for companies and universities, as they have limited storage capacity. Early-stage prototypes are usually rough and of low-fidelity, and are therefore often discarded or substantially modified during the course of a project. This is a problem for researchers wanting to investigate the use of such early-stage prototypes, as it means that getting accurate temporal data from retrospective studies is difficult due to the availability of the prototypes and the resources available to the researchers.

One way of collecting data on prototypes is for the researcher to physically collect (and store) every prototype of a project, either during or after the project has finished. If prototypes are physically collected during the project, this hinders the project collaborators from modifying or building on existing prototypes. Conversely, if prototypes are physically collected after the project has finished, only the last iteration of a given prototype is collected, as discussed by Sjöman et al. (2017).

## 1.3 Problem statement and Research Objective

The authors have identified the combination of few observations and extensive use of student participants in studies using activity-focused research methods as a symptom of lacking tools and methods for capturing (and therefore researching) prototyping as a design activity in engineering design. Current output-focused research methods either cause interruptions of the design process by removing the prototypes from the project contexts or relies on availability of and access to them after the projects have finished. Many of the studies on prototyping activity and output use established methods that are laborious and resource intensive for collecting and analysing data, resulting in low number of observations (i.e. data points) per study—especially when gathering data from industry (Erichsen et al., 2019).

Consequently, there is a need for new tools and methods that can capture more observations on prototyping (the activity) and prototypes (the output from design activity), yet still offer enough level-of-detail on iterations made during the development projects so that researchers can use this data to choose where to apply the existing, more resource demanding methods. Beyond the mentioned incentives for capturing prototypes for engineering design research, a prerequisite for getting the presented research tool for capturing prototypes adopted by other researchers is that existing analysis methods can be employed to analyse the data, and also that the data can be analysed by other engineering design researchers.

Essentially, the core problem addressed in this article is that existing research methods and tools prevent researchers from capturing and analysing enough observations on design activity and its output. Based on this problem statement, the authors' research objective is to provide a research tool for capturing more observations from development projects, and to do so by focusing on capturing prototypes (as output from design activities).

This article presents the foundations of this research tool for capturing prototypes from ongoing projects, it's minimum requirements, system specifications and implementation, as well as an evaluation of the tool used for collecting and analysing prototypes from both industry and academic development projects over a period of 17 months. During this period, the system has had 76 individual users who have contributed to a total of over 800 (850 as of March $20^{th}$, 2019) captured prototypes.

## 2 Design Objectives for Capturing Prototypes as a Proxy for Capturing Activity

A set of design objectives has been used to guide the various design decisions when developing the research tool presented in this article. A summary of the design objectives can be found in Table 1.

Table 1 Design Objectives

| Obj. # | Design Objectives |
|---|---|
| 1 | *Collect data with minimal intervention by the researcher.* |
| 2 | *Capture the prototypes once they are created, with minimal interruption of the designers' workflow.* |
| 3 | *Add value to the designers' documentation workflow to incentivise use of the capture system.* |
| 4 | *Create an expandable and scalable system that can be scaled to multiple locations and users.* |
| 5 | *Capture the following dimensions from the prototype itself;*<br>- *physical appearance and form*<br>- *designer's intent / driving question(s)*<br>- *who participated in building the prototype*<br>- *when the prototype was built* |

Objective 1: *Collect data with minimal intervention by the researcher.*

Ideally, the research tool should be able to capture all prototypes of every project that the researcher wants to study. One way of ensuring that as many prototypes are captured as possible is for the researchers to oversee that all prototypes that are created during the project are also collected (either physically or digitally) by the researchers, similar to what is done by V. Viswanathan et al. (2014) and V. K. Viswanathan & Linsey (2012). Although practise may work for small-scale research following only a limited number of projects, e.g. a university course, this approach would not be feasible for gathering larger amounts of data, e.g. multiple university courses in different continents or design teams in multiple companies, as discussed by (Hofstede, 1984). Additionally, researcher intervention of the design process may alter project outcomes leading to capturing less applicable data.

To solve the resource problem of the researcher not being able to supervise each project rigorously, and to minimize the amount of researcher intervention required for capturing prototypes, this article advocates that *digital* capture of physical prototypes should be based on self-reporting—relying on the designers to capture their own prototypes as they are created during the projects. Effectively, this means that the prototypes can be captured without requiring the researcher to physically interact with any of the prototypes in order to capture them.

Objective 2: *Capture the prototypes once they are created, with minimal interruption of the designers' workflow.*

A consequence to capturing prototypes digitally is the possibility to capture the same artefact at multiple instances as it is modified throughout the project. In order to capture such occurrences, there is an emphasis on capturing the prototypes as soon as possible after they are created. The emphasis is on capturing prototypes from *ongoing* projects, as opposed to capturing the prototypes retrospectively, i.e. after the projects are finished. It is important to capture the changes made to the artefacts—i.e. how the projects progress through various iterations and design decisions, to ensure that valuable changes to the artefacts are not missed or ignored.

While digitally capturing physical prototypes from ongoing projects solves the practical challenge of having to remove the prototype from its environment (i.e. the project and its participants), it can still cause interruptions of the designers' workflow. Minimizing such interruptions (e.g. having to pause the design activity in order to capture a prototype) is important to ensure that iterations are not skipped or ignored. For the data to truly represent all the prototypes of a project, all prototypes created throughout the entire project should be captured so that no prototypes are 'missing' (i.e. not captured).

Objective 3: *Add value to the designers' documentation workflow to incentivise use of the capture system.*

In order to gather more observations (i.e. prototypes), and because this form of data capture should be voluntary, the implementation of this tool needs to be user-centred (Abras, Maloney-Krichmar, & Preece, 2004). This means that capturing prototypes (through self-reporting, as previously discussed) should be truly effortless, requiring as little effort from the designers as possible. As discussed by Sjöman et al. (2017), testing shows that lower effort required to capture prototypes yield higher number of captures (and consequently, lower number of 'missed' or 'ignored' prototypes) with the same amount of prototypes in a project. Notably, there is a trade-off between usability and capture fidelity (i.e. level-of-detail). Findings by Sjöman et al. (2017) also imply that the capturing should be done near the designers' workspace, as having to bring prototypes to a remote location for capturing prototypes would increase the required effort considerably, and thus increase the amount of prototypes not captured by the research tool.

To further incentivise capture of prototypes (for the project team), there needs to be some other form of added value for each person capturing data. For the designers, access to captured prototypes provides a useful foundation that can be used documentation, which is often required by many companies. Consequently, having access to the captured prototypes remotely also increases the value, both in documenting and enabling sharing to other colleagues and collaborators.

Objective 4: *Create an expandable and scalable system that can be scaled to multiple locations and users.*

A benefit of digitally capturing prototypes based on self-reporting is that it enables gathering data from multiple users and locations simultaneously. To leverage this, it is an objective to design the research tool in such a way that enables capturing prototypes at multiple locations by different users, and that all captured prototypes are stored in the same repository.

Objective 5: *Capture the following dimensions from the prototype itself;*

- *physical appearance and form*
- *designer's intent / driving question(s)*
- *who participated in building the prototype*
- *when the prototype was built*

The last objective considers what to capture from the prototypes. From inspecting a single (physical) prototype, there are the explicit, physical properties to consider. Physical properties include what geometry (i.e. shape and size) the prototype has, along with weight and specific material-related properties, e.g. density, surface texture, conductivity and reflectivity. It's worth noting that some of these material related properties are also a result of tooling (M. B. Jensen, Balters, & Steinert, 2015), e.g. surface texture from sanding, machining or Fused Filament Fabrication (FFF) 3D-printing. Beyond the physical properties, prototypes also inhibit other, less explicit (i.e. 'tacit') attributes (Auflem et al., 2019). Essentially, these attributes answer questions regarding the 'why', 'how', 'who' and 'when' of the prototypes.

The 'why' of a prototype is often referred to as purpose (or intent)—the reason for building the prototype (M. B. Jensen et al., 2015; Lim, Stolterman, & Tenenberg, 2008). Lauff, Kotys-Schwartz, & Rentschler (2018) highlight three different roles of prototypes in companies; to facilitate communication, to support decisions and to aid in learning. A prototype may have multiple purposes, and these dictate attributes like functionality, interactivity and fidelity (M. B. Jensen et al., 2015; Lim et al., 2008). Therefore, the authors argue that capturing the designers' intent of the prototype—e.g. if a prototype was intended for learning or for verification/testing—is important to understand why and how the prototype was made (Sjöman et al., 2017) and what was learned from prototyping. In order to contextualise such information, it is also important to capture the creators of the prototype – to know whose intents and purposes are being captured.

Schrage (1993) states that a crucial ability for any organisation is to learn from its own projects, and having access to multiple prototypes across different projects over time can reveal patterns (or lack thereof) that can tell whether the organisation is learning. Schrage (1993) claims that "by comparing the prototype changes per cycle, management now has a rigorous vehicle for measuring progress", claiming that prototypes may be a key to measuring project progression, especially when developing new radical product innovations. Schein (1990, 2004) suggests inspecting an organization's culture through artefacts, somewhat similar to Schrage (1993), who states that prototypes are the keys to understanding a organization's culture.

Prototypes provide an interesting opportunity to investigate the temporal dimension of a project – being tangible evidence of the project's directions as they evolve over time. Another interesting application of having temporal data on prototyping is to see if prototypes are reused, repurposed or modified throughout projects, indicating that prototypes themselves may be temporal, as well. Such reuse could be due to design fixation (V. Viswanathan, Tomko, & Linsey, 2016), or simply to save time and resources. In summary, capturing temporal data of the prototypes allows for new research topics that could be investigated further.

To summarize, this article argues that the four most important dimensions to capture from prototypes (when using captured prototypes for research) are the physical properties of the prototype (especially the shape, appearance and materials of the prototypes), as well as the time the prototype was created, the creator(s) of the prototype, and the intent of making the prototype. Additional information is both beneficial and relevant to the researcher, but detailed contextual information of the prototype can also be collected retrospectively by leveraging the initial, minimum viable information that is captured—given that the researchers have access to the project team who is creating the prototypes.

## 3   System Design

A cyber-physical system for digitally capturing physical prototypes has been created by the authors, shown in Figure 1. In this system, the prototypes are captured through multi-view images, along with metadata describing by who, why, when and where the prototypes were captured. There are other ways of capturing, and the rationale behind generating multi-view images was based on user feedback from earlier iterations (Sjöman et al., 2017) and provided a starting point for how to capture the prototypes. Figure 2 shows how the proposed system could be implemented for capturing prototypes from ongoing projects. The system described here is used to capture the data shown later in the article.

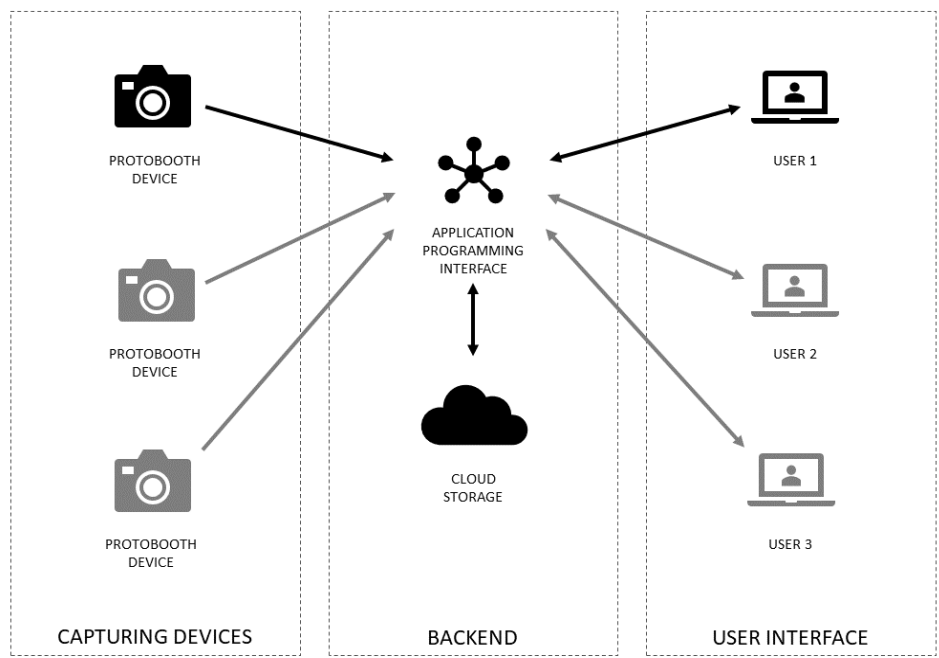

Figure 1 Schematic of the capturing system's main components.

The capturing system is made of three main components:

- A physical capturing device for capturing the data
- An online and cloud-based backend for handling and storing the data
- A user interface for interacting with and enriching the data (as users, both designers and researchers)

The multi-view images and metadata are generated by a connected installation (often referred to as a 'photo booth for prototypes' or 'Protobooth' by its users), and the content is handled and stored in a cloud-based storage service system. The users capture prototypes with the swipe of their workplace (RFID) access card (which is unique to each user), and the system has a web-based interface for studying the captured data. In this web-based interface, users can organise scanned content by projects, combine content scanned by multiple users, as well as add titles and descriptions to each captured prototype and project, thus enriching the metadata. Although there is one physical capturing device shown in this implementation, there could easily be multiple capturing devices at different locations capturing prototypes and sending the captured data to the same backend, as shown in Figure 1.

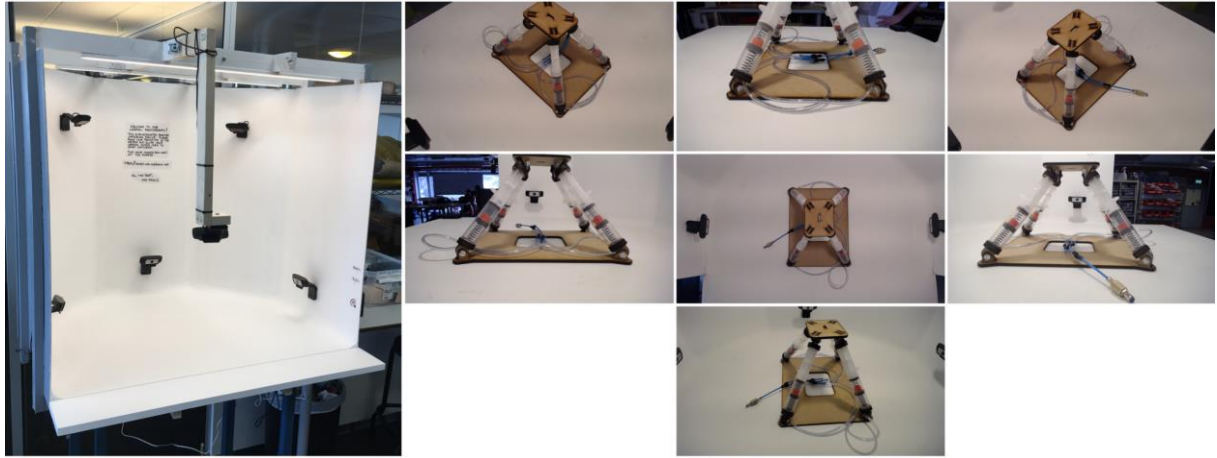

Figure 2 Left; the described physical capturing device used to gather data, a 'photo booth for prototypes', and right; example of multi-view images of a single prototype ('Prototype 37' from the example project) generated from the capture system.

A physical capturing device was chosen over a mobile application-based capturing system to ensure that all users would be able to access the system regardless of personal mobile devices. A lot of the designers use personal mobile devices with different hardware specifications and operating systems, and a physical capturing device was chosen to ensure that the captured data would be of sufficient quality and to streamline the capturing process (i.e. the effort required for capturing a prototype). Moreover, having to interact with personal mobile devices for capturing the prototypes may further interrupt the focus and attention of the designers, interrupting the design process.

The physical 'Protobooth' for generating the multi-view images is roughly one cubic metre in volume and is painted white and has strong overhead diffused lighting, shown in Figure 2. The booth is powered by a small desktop computer running Linux operating system, and has an online connection for uploading the captured content. There are seven (7) webcams with Full-HD (1920x1080) resolution mounted at various angles, all facing inwards towards the centre of the booth. The camera angles of the multi-view images are dubbed 'front', 'top', 'right', 'left', 'rear right', 'rear left' and 'rear' and the output from these seven angles is shown in Figure 2. In addition to the cameras, the booth has a physical RFID reader for reading user input and an Arduino for managing two status indicator LEDs. The system detailed in this article features a cube-shaped white backdrop and uses seven webcams for taking multi-view images. This cube-shaped backdrop is suitable for prototypes of up to approximately 40cm by 40cm by 40cm shape.

The system is powered on by default and is activated with the swipe of an RFID card in close proximity to the RFID scanner. The user is instructed to place the prototype inside the physical booth before activating the system through the RFID scanner. Upon activation, the computer runs a series of scripts taking a photo for each of the seven webcams and uploading this to the system's backend. Additional metadata to this upload includes information about when the prototype was captured (i.e. through a UNIX timestamp), as well as which RFID card and physical booth were used. As previously stated, one core philosophy of this system is that designers are able to capture their prototypes with as little effort as possible. Consequently, an RFID-based single-operation capture sequence has been chosen. Capturing the multi-view images of one prototype takes approximately 9 seconds, and the user is notified through the status indicator LEDs when the capturing is done.

The capturing system has a backend for handling and storing the information captured by the physical booth. This backend consists of a NodeJS Application Programmable Interface (API) for handling the various flows of data, and a cloud-based storage service for storing the data. This storage system handles the multi-view images and the metadata separately for speed, security and redundancy reasons, storing the metadata in a NoSQL database and the images in a separate Object Storage database.

The capturing system has two general ways of interacting with the captured data. Primarily, the data is accessed through a web-based interface that is programmed using ReactJS ('React – A JavaScript library for building user interfaces', 2019) and Express ('Express - Node.js web application framework', 2019). In this interface, users can organise their own content into various projects and annotate the prototypes (e.g. by adding the purpose/intent, testing procedures or other valuable insights linked to that particular prototype). Each project can have multiple contributors, meaning that one project can have prototypes captured by different users. The users are also able to add more data to each prototype, e.g. title and description, thus adding additional information to the metadata. Additionally, the backend can also be used for accessing the 'raw data', e.g. for analysis or debugging, which is done through specific API calls.

## 4 Results from Testing the Proposed Tool for Capturing Prototypes

To test the presented capture system, the authors have implemented the capture system in a prototyping laboratory dedicated to early-stage engineering design research at the Norwegian University of Science and Technology (NTNU), where it has been adopted by the laboratory's users. This prototyping laboratory uses real-world challenges from industry as setting for studying early-stage engineering design tools and strategies. As the authors have daily access to this laboratory, access and monitoring of the system has allowed doing several iterations in order to reach the fidelity and complexity presented in this article. Testing of the capture system has also included collaboration and pilot-testing at a R&D department in a market-leading producer of medical simulators, as well as doing in-house pilot testing at the aforementioned prototyping laboratory at NTNU.

The capture system has been used for collecting data in this environment for 17 months. During this period, the system has had 76 individual users who have contributed to a total of over 800 (850 as of March 20[th], 2019) captured prototypes. Figure 3 shows the distribution of prototype captures by time of the day. In this figure, the colours are included to indicate to which project the capture belongs, and this figure shows that the capture system is able to capture when and by whom a prototype was captured. The vertical spread (i.e. 'jitter') for each date is a random value applied to each dot in an attempt to separate the dots from each other to avoid over-plotting. As the capture system is still in daily use, the amount of content continues to grow steadily over time.

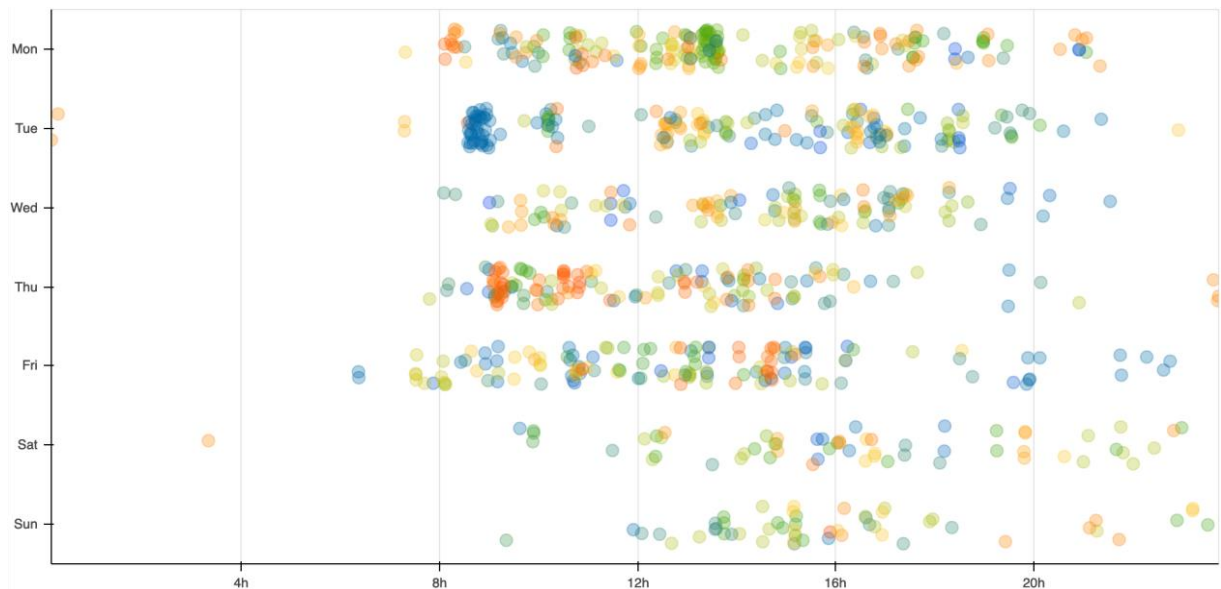

Figure 3 Protobooth uploads by time of day (horizontal axis), sorted by weekday (vertical axis), with colours indicating different projects.

## 5   In-Depth Analysis of the Captured Output of a Single Project Case

One project case is analysed in detail to demonstrate how data captured by this system can be used for researching early-stage engineering design projects that focus on prototyping for concept generation. This includes a review using of multi-view images as well as analysis of the metadata that is automatically generated by the capture system. Analysis of this metadata provides insight into project progression (i.e. how the project develops over time), including the frequency of prototyping, which days the project participants are most active, and how the prototyping changes.

The project case highlighted in this article is the development of a new Cardiopulmonary Resuscitation (CPR) mannequin for a medical simulator context. The project has one participant, a graduate student at the Norwegian University of Science and Technology (NTNU), developing a concept for simulating CPR. The project was used as a basis for the student's master's thesis. The development challenge was set in collaboration with a leading manufacturer of medical simulators and was intended as an early-stage development project without particular design requirements or specifications. The challenge was formulated as "re-thinking CPR mannequins for medical simulations". CPR is a complex topic, but can in short be described as series of short forceful compressions to the patient's chest, followed by assisted respiration (through blowing air into the patient's lungs) in order to resuscitate the patient. Compression parameters like rhythm, force, depth and hand positioning are key to effective CPR. In CPR simulations, the patient is often replaced by a simulator device, e.g. a mannequin that resembles a human person.

Being part of the prototyping laboratory dedicated to early-stage engineering design research at NTNU, the student had access to several tools and machinery over the course of the project, e.g. laser cutter, hand tools, vacuum former, 3D-printers, CNC-machining equipment etc. In order to record the data presented in this paper, the student was asked to capture the prototypes produced throughout the project through the previously described capturing system, and to capture prototypes as they were made. The student was asked to make his own judgements on how to define iteration and revision. The student had the following remarks on how and why he decided to capture each prototype:

*"I captured the prototypes when I was satisfied with the design and when each prototype was able to fulfil the purpose of why it was made. If I made modifications to a prototype, I also captured the modified design. Usually, the purpose was to test something for myself or to interact with users, or a combination of both. It was also important for me to capture the things that did not work, both so I could remember what I had done and so I could see what didn't work as well as what worked." – Student interviewed on May 15th 2018*

Over the course of the project, from October 2017 through May 2018, the student captured 82 prototypes. A more in-depth discussion on some of the prototyping technologies and principles used in this particular project is done by Auflem et al. (2019). A timeline showing the prototypes created during the project is shown in Figure 3, indicating that the prototypes were captured in two main periods; one in October through November 2017 and another in January through May 2018. The large gap visible between November 2017 and January 2018 is due to the student having exam period followed by Christmas holidays. A table of the 82 captured prototypes is included in Appendix A, sorted chronologically from earliest to latest time-of-capture.

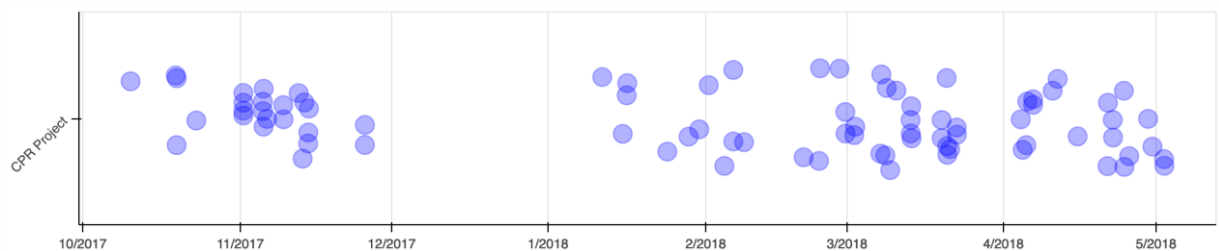

Figure 4 Timeline of the presented project case, made using automatically generated metadata of captured prototypes from October 2017 through May 2018.

Figure 4 demonstrates that the capture system automatically creates a record of the prototypes produced throughout the project with accurate timestamps. The horizontal axis shows time of capture. The vertical spread is applied for visualisation purposes to avoid over-plotting. As all dots have the same opacity, darker colours indicate a cluster of prototypes within a shorter period. Each of the dots in this timeline represents a prototype captured through multi-view images and corresponding metadata. For each of the dots in the figure, there are multi-view images and metadata that can be used to further elicit information about the project, e.g. the multi-view images shown in Figure 2 shows the 'prototype number 37' from this example project. Having access to this timeline information, in addition to the multi-view images of the captured prototypes gives the researcher detailed information about what happened when.

## 5.1 Manual categorization of materials, tools and disciplines used to create the prototypes

In addition to using the data that is automatically generated from each prototype capture, it is possible to use manual inspection of the multi-view images and categorise the prototypes. The rationale behind doing such a manual categorisation is to investigate if the captured data can be used for analysis of the projects' prototypes. This data can be categorised using various coding schemes. In this paper, three different schemes for coding the prototypes has been chosen to exemplify the practical application of the captured output from the presented system:

- The material used in each prototype
- The tools used to produce each prototype
- The disciplines required to produce each prototype

The three categorisation schemes have been applied through visual inspection of the multi-view images of each prototype. There are two aspects that are interesting when applying such categorizations to the prototypes. The first is to generate descriptive statistics on material and tool usage of a project, giving insights into trends and habits of the project team. The second is to show the development and progression over time, showing the specific materials, tools and disciplines used at the various iterations and events throughout the project. The latter will be discussed further in Section 4.3. One person categorised the 82 prototypes using the three different schemes. Since this categorization is used to exemplify how to use the data generated by the capturing system, inter-coder reliability has not been tested for the three categorisation schemes. However, since the multi-view images used as input for the assessment are available, this categorisation could be repeated by several coders if necessary.

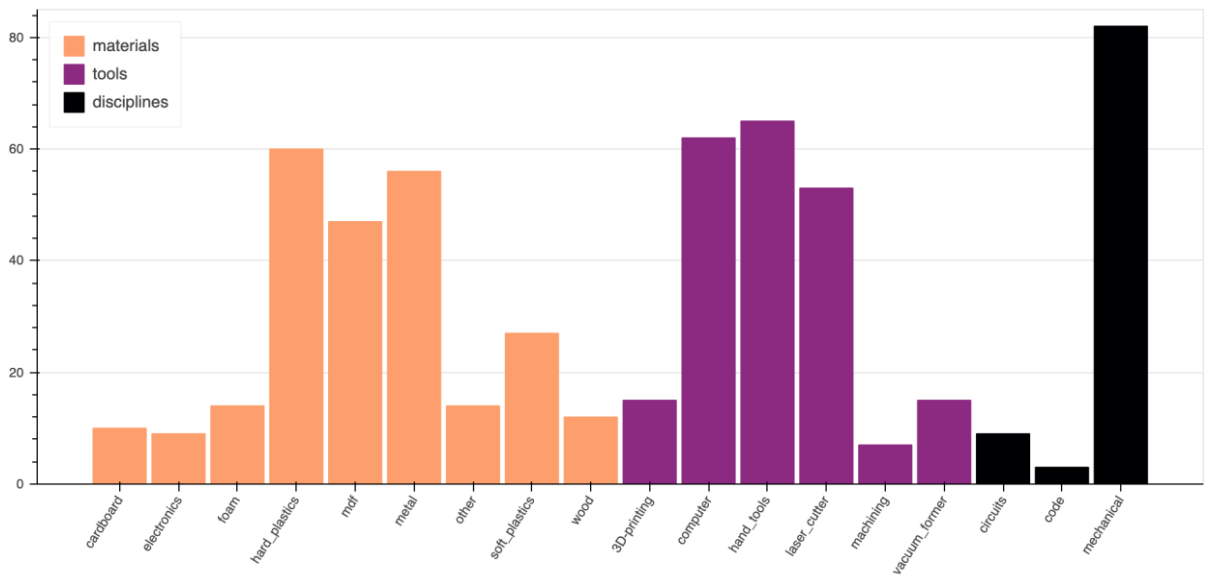

Figure 5 Cumulative summation plot describing the total number of materials, tools and disciplines used throughout the project.

A project overview can be found in Figure 5, showing a summary of all the materials, tools and disciplines used throughout this project. For categorising materials used in the prototype, the materials registered are only those that are present in the multi-view images. If a material was used to make the prototype, but is not included when capturing the prototype, then it is not included in this coding scheme. The 9 material categories are foam, cardboard, medium-density fibreboard (MDF), wood, hard plastics, soft plastics (including rubber), metal, electronics, and other. This categorization is somewhat coarse and does not differentiate pre-made parts (e.g. nuts and bolts) from stock materials (e.g. sheet metal) but serves as a basis for showing that different material coding schemes can be applied.

Six tool categories were created from the most common tools found in the participant's regular prototyping workspace to categorise what tools was used to produce each prototype; hand tools (e.g. saw, hammer, screwdriver, ratchet, spanner, vice, etc.), 3D-printer, laser cutter, machining (e.g. milling or lathing), vacuum former and computer (e.g. CAD-work or programming). The prototypes were labelled in these categories through visual inspection of the multi-view images, and only includes the tools used to make the prototype that is captured—e.g. if a captured prototype includes a 3d-printed part, it has been assumed that a 3d-printer has been used to make the part. Likewise, if a captured prototype includes sheets of MDF with visible soot on its cut edges, it has been assumed that the part has been cut by a laser cutter. Again, being a coarse categorization, this serves as an example showing some early-stage prototyping tools logged during a development project. For assessing what disciplines were incorporated into each prototype, the prototype was labelled in three categories; mechanics (i.e. the prototype includes physical (moving) parts), software (i.e. the prototype includes programming of some sort, e.g. includes a microcontroller) and electronics (i.e. the prototype includes circuitry, e.g. wires and solder).

To show the project's development and progression over time, a plot of the various materials found in the 82 prototypes is shown in Figure 6 and Figure 7. In these two figures, the prototypes are sorted chronologically from left to right.

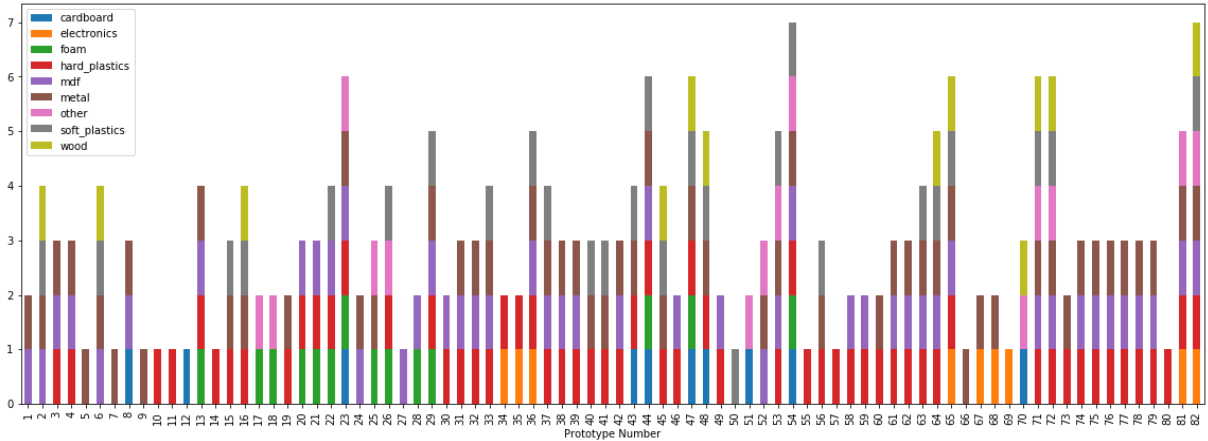

Figure 6 Materials per prototype, sorted chronologically from left to right, with each bar representing a prototype.

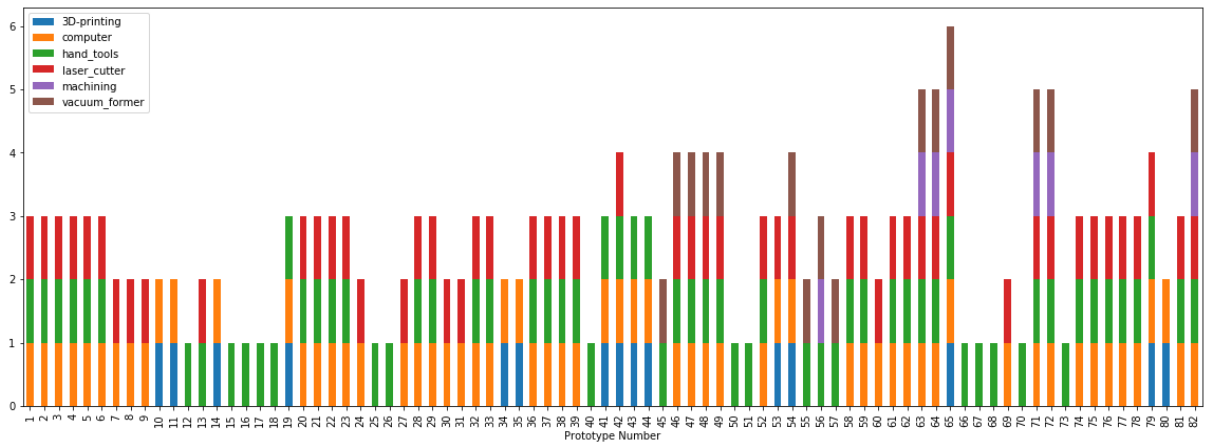

Figure 7 Tools per prototype, sorted chronologically from left to right, with each bar representing a prototype.

### 5.2  Manual coding of solution principles

Another project-specific assessment of the prototypes is to categorise the prototypes by solution principle. The aim of coding the prototypes by solution principle is to show how projects can be retrospectively assessed using data from digitally captured prototypes. There are multiple ways of doing this, and the authors have chosen to manually code the prototypes by what solution principles are embedded in the prototypes through both visual inspection and by talking to the designer and discussing the various steps taken throughout the project.

This project includes three main categories of solutions for making the CPR mannequin; the compression of the chest, the breaking of ribs and the logging of data (in order to log user training performance). Each of these three categories have several sub-categories, e.g. compression by using a spring. A plot of these solution principles, sorted chronologically from left to right, is displayed in Figure 8.

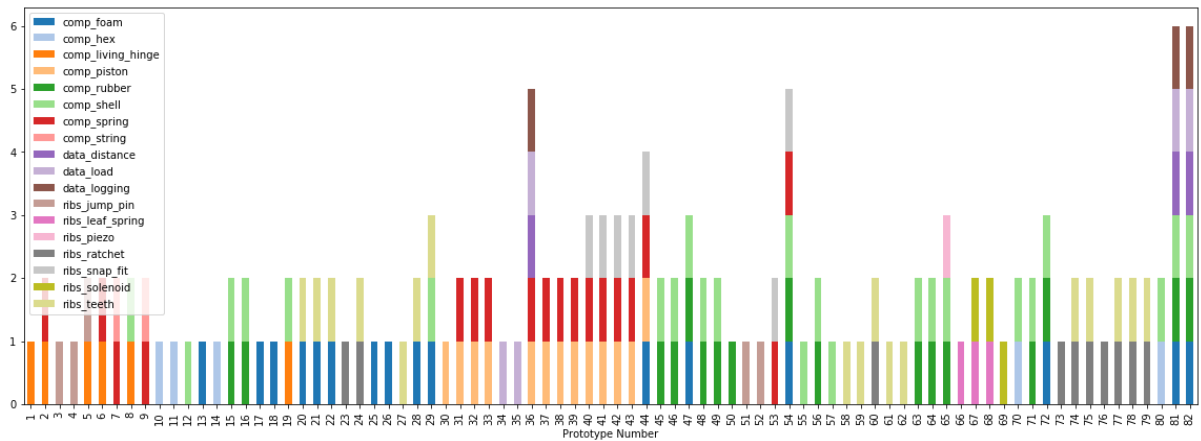

Figure 8 Solution principles per prototype, sorted chronologically from left to right, with each bar representing a prototype.

## 5.3 Observations from Investigating the Example Project Case

The data shown in the previous section show that it is possible to get detailed quantitative measures with (relatively) little effort when investigating prototyping activity in captured project cases. Looking further to exemplify the use of this proposed research tool of capturing prototypes, this section presents some key findings from this one project case and shows these key findings through the data.

One of the most interesting aspects of collecting all the prototypes of a given project is the ability to visualise the relationships between the prototypes. In Figure 9, such a visualisation is presented, coded by the student who also made each of the prototypes. Each node in the figure represents a single prototype; the nodes represent physical prototypes used internally within the project, the blue nodes (i.e. Prototypes 5, 17, 29, 60 and 63) represent prototypes tested externally (e.g. to receive feedback from users) and the green node represents the final concept (i.e. 'Prototype 82'). The prototypes are sorted chronologically along the horizontal axis, and the vertical spread is applied to avoid over-plotting. This figure somewhat resembles the notion of 'linkography' (Goldschmidt, 2016), yet in this specific example is applied to visualise links between physical prototypes. Images of each of the 82 prototypes are included in Appendix A.

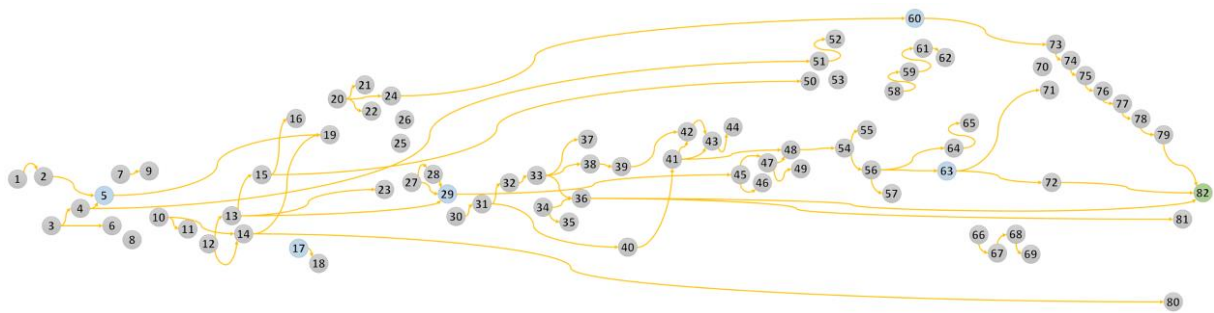

Figure 9 Links between the 82 prototypes of the in the example case project, sorted chronologically from the earliest (i.e. "Prototype 1") to the latest, final concept (i.e. "Prototype 82").

Essentially, Figure 9 shows the various routes taken to reach the final concept (i.e. "Prototype 82"). Though there are several 'disconnected' nodes in this figure, that are not directly linked to the final concept, they provide interesting starting points for further investigation by researchers.

One interesting observation on material usage during this project is the varying amount of materials per prototype throughout the iterations. By investigating the increased amounts of material per prototype, it appears that there is a strong indication that this occurs whenever the designer is refining the prototype designs, adding features or parts to create more complex prototypes. This is apparent in Figure 6, where there is a gradual increase in prototyping fidelity from 'Prototype 20' through 'Prototype 22'—increasing the amount of material types per prototype through the iterations. One can argue that these prototypes share a lot of similarities in-between themselves, and that they can be interpreted as iterations on the same principles rather than stochastic or 'independent' prototypes—also apparent in Figure 9. Similarly, the sequence of prototypes from 'Prototype 30' through 'Prototype 43' indicates that the project participant had found a promising set of solution principles. Here, focus shifted towards elaborating similar concepts, and there is a clear indication of fidelity and complexity increase per prototype.

During the project, the student had multiple sessions sharing ideas with the collaborating company. After one such session, where the student presented the current project focus to the company, there was a noticeable shift in the direction of the project. This is visible in the data as an abrupt change in both number of materials (Figure 6) and complexity (Figure 8) of the prototypes produced (after 'prototype number 43') and could be used as an entry point for further investigation. It is worth noting that finding this event is possible through visual inspection of the data. However, finding more details on the cause(s) of the event is substantially more difficult, and requires some form of interaction with the project participant.

Another observation worth mentioning is the alternating of focus on various solutions during the project. In trying to create a new concept for simulating CPR, the student has focused on recreating two main aspects; creating a realistic compression mechanism and to simulate the breaking of the patient's ribs. When inspecting Figure 8 and Figure 9 simultaneously, it is apparent that the focus of the project is not linear – rather a continuous effort along parallel 'threads'. The concepts are alternating between breaking ribs and compression mechanisms, shown in Figure 8—sometimes at the same time and sometimes individually. This observation is helpful to show various concepts that designers consider (and test) during a project.

## 6 Discussion

The aim of the presented research tool is to enable researchers with more data on engineering design cases. This is useful for getting empirical data in the first place, but also for giving multiple researchers the ability to research the same projects, hopefully also leading to increased quality of engineering design research. The capture of more than 800 prototypes has shown that this research tool may be an asset for engineering design researchers wanting to study output from design activity.

Digital capturing of prototypes enables two approaches for enriching case studies of a given early-stage engineering design project. The first approach is to use the captured prototypes (exemplified through multi-view images) for a qualitative assessment of the projects, which can provide new insights and understanding on various aspects like design decisions, trade-offs and specifications e.g. when coupled with interviews. The second approach is to analyse the metadata provided by the system to give understanding into prototyping patterns in the projects, which e.g. can be used as proxies for project progression. This article has demonstrated both these approaches with examples from real-world projects. As the capture system enables continuous capture of physical prototypes, it provides a countermeasure related to retrospective engineering design case study techniques, such as interviews or protocol studies.

This article argues that there are four dimensions that are especially important to capture when using the data for engineering design research; the physical properties of the prototype (especially the shape, appearance and materials of the prototypes), as well as information about why, when and by whom the prototype was made. As discussed earlier, there are benefits to capturing more information than what is stated above, but the four said dimensions are considered especially important. It is possible to argue that there are types of prototypes that are more important to capture than others, yet capturing pre-defined and pre-labelled categories would go against the aim of the proposed system—namely to provide a quantifiable (and therefore more objective) capture of physical prototypes.

Since this research tool is relatively new, it is worth noting that there might be other possible implementations that are suitable. The previous sections have shown that the capture system is able to capture extensive information about the captured prototypes that can be used to perform manual coding, similar to what can be done with existing tools and methods. Moreover, the testing of the capture system described here also shows that a low-effort capture of prototypes leads to more empirical engineering design project data, which leads to the data collection process being more scalable compared to existing tools and methods, enabling larger data sets to be studied. This is a clear improvement from current practise, where there are few studies that extensively use systematically captured prototypes (Erichsen et al., 2019).

Some of the features of the described capture system are context specific, and the specified system is a result of a prototype-driven exploration. There has for example been a stated need for multiple views of the prototypes, as assessing the prototype's function and materials from one angle have in some cases been difficult. This redundancy has little impact on the way the system is implemented, yet has opened up for future exploration, such as creating 3D models from multiple photos (i.e. 'photogrammetry') (Kohtala, Erichsen, Sjöman, & Steinert, 2018).

Since one of the design objectives have been to minimize the interruption of the designers' workflow, the low-effort capture is done at the expense of (initially) capturing more detailed metadata. E.g., by using RFID access card as a proxy for who captured the prototype, it is assumed that this person also was part of creating that same prototype. Detailed metadata can be added later, e.g. by using a web-based user interface as previously described. Filling the metadata does not happen automatically, but in a case of research, the researcher has an incentive to fill the blanks of the metadata. This was the original intention of the project as well, but it seems the project has not yet found the means to incentivise the end-users to fill in the metadata by themselves. To summarise, the authors do indeed deem it feasible to gather the four dimensions of interest, yet much work needs to be done in order to (effortlessly) gather more contextual information beyond these four dimensions.

# 7 Strengths and Limitations

An observation is that it would have been difficult to collect this amount of accurate data (i.e. more than 800 prototypes) without the proposed capture system, and it would have required a large amount of effort by the researchers. Likewise, it would have been hard to get the contextual information without interaction with the student making the prototypes. This reflects that the proposed tool is a valuable supplement to existing research tools, rather than a 'set and forget' automated capturing device.

In J. F. Erichsen et al. (2019), the extensive use of student participants (as opposed to professionals) in engineering design research is pointed out as a potential shortcoming in current literature. This article argues that it is preferable to gather data from both industry and student projects in order to compare the contexts and to evaluate the validity of the student projects used as proxies for 'real' engineering design projects. The authors recognize that using university courses for iterating on methods and experiments setup is useful, since the data is more accessible and controllable than collecting data from industry. The authors suggest gathering larger data sets from both industry and student populations to ensure high ecological validity of the studies. This is exhaustive with today's tools, and not very feasible for a single research group, but would be considerably easier with the tool proposed in this article.

Since the proposed tool is in its premature stages, the authors would like to point out that the system is a research tool and is not an automated documentation tool for prototypes. The focus of this tool is the capture of prototypes and supplementary metadata for research. The authors recognise that the output can later be used for documenting the process, although this is not the primary focus of this article.

There are limitations to the assumptions that designers capture the prototypes that they make and that the designers capture the prototypes at the time that they make them. With a system for capturing prototypes, it is difficult to determine what has not been captured without a regular presence in the areas where the prototyping takes place, which is a clear limitation to this type of data collection. Moreover, digitally captured prototypes might lack information on external factors that impact on the project's development. Hence, using this tool without doing some form of interviews or follow-ups might lead to missing out on some key insights. During testing, it was found that users need some time to start using a capture system. However, once familiarised with the system, it is observed that most of the designers use the system for capturing all their new prototype iterations.

Through testing, it was found that one of the 76 individual users preferred to capture prototypes in 'bulk', scanning many (e.g. more than 20) of the prototypes they made, and doing so in a short amount of time. This by itself does not invalidate the captured data, but it does make any automated timestamping of that particular user's prototyping data inaccurate—though this should not be mistaken for (programmatical) 'errors'. This can easily be rectified with editing dates manually per prototype capture, given that the inaccuracy is picked up and noticed by the researcher. Through testing, we have found these 'bulk captures' to stand out in the data, making them quite easy to spot though various time-related visualisations, e.g. the three clusters of colours visible in Figure 3.

Another challenge of having designers capturing their own prototypes is determining what (and who) defines an iteration (i.e. how to separate one prototype from another, or how to decide when a prototype has been substantially modified to call it a new iteration). During the nine-month testing period, the authors have not instructed the participants on what defines an iteration or how to decide when a prototype is modified 'enough' to deserve being captured a second time. Interestingly, on visual inspection of the data, it is found that almost all sequential prototypes in the various projects have substantial edits made to them. The authors do stress that one cannot say anything on the prototypes that may not have been captured, but that this tool provides an interesting opportunity to research how designers define iterations.

## 8 Future Outlook

This research is considered to be continuously developing, and there are several points that should be highlighted for future investigations, as well as new research questions emerging from investigating the captured prototypes. The authors encourage other researchers to experiment with numerous ways of capturing prototypes. As previously stated, the capture system described in this article is an example used to show the proposed method but is not the only solution to capturing prototypes. For example, the authors are experimenting with 3D-scanning and reconstruction through photogrammetry (i.e. creating 3D-geometry from pictures, as shown by Kohtala et al. (2018)), yet there is still some way to go before such features are ready for full-scale testing. The authors are also experimenting with the possibility of adding other sensors (e.g. load cells, microphones, etc.) to further enrich the metadata (Kohtala et al., 2018). The authors intend to further test various other solutions to capturing prototypes. There are early indications from pilot testing that weight-measurements (through load cells) can be integrated and added as a metadata property. Various projects have varying prototype capture needs and challenges, and size and format of the data is possible to tune to specific contexts. For example, someone prototyping mechatronic systems might want a smaller (and possibly higher resolution) capture of electronics and printed circuit boards for capturing prototypes, whereas others might want a bigger setup—i.e. for capturing physically larger prototypes. The exemplified tool for capturing prototypes can be tailored to fit specific contexts, as different companies and facilities need different capabilities.

Though the system presented in this article is mainly aimed at analysing the captured data without interfering with the projects or project participants, it would also be interesting to explore how such a capturing system could be used to enhance or guide the design process. There is research targeting how intervention in design activity might affect outcome, as described by (Tang & Leifer, 1991; Törlind, 2007; Törlind et al., 2009), and though most of said research is focusing on intervening based on observed activity, this capturing system could potentially employed to enable intervention of project progression between design sessions and activities.

Since the example capture system shown in this article is using multi-view images for capturing prototypes, the authors are experimenting with possibly classifying parts of this output automatically. With digitally capturing large quantities of data, the presented research tool presents the foundations for being able to train various artificial intelligence-based predictors and classifiers that can be used in engineering design research. Therefore, while the categorisation of materials, tools and disciplines are applied manually in this article for the sake of exemplifying the use-case, the authors deem it feasible (based on preliminary tests) to use the multi-view images for automatic categorization (e.g. categorizing materials using a pre-trained convolutional neural net trained for image classification).

## 9  Conclusions

In this article, a research tool for capturing prototypes with the aim of enabling more and better empirical research of early-stage engineering design projects has been proposed. This proposed tool has been exemplified through a cyber-physical capture system, showing that it is feasible to capture empirical data from early stage engineering design projects, and that it is feasible to use captured prototypes as input for studying progression in these projects. The data output from this capture system has been used to provide quantified data on prototypes as a proxy for design activity, and examples of enriching a case study on an early-stage engineering design project have been presented. Although mainly aimed at enabling research of early-stage engineering design projects, it is hypothesised that there are many other stakeholders that could be interested in quantified data on prototyping activity from engineering projects. This includes (but is not limited to) engineering design researchers, HR teams, project managers, prototyping workshop facilitators and lastly the designers themselves.

With proposing a new tool for capturing prototypes, the authors aim to enable capturing larger engineering design project data sets so that early-stage prototyping can be analysed extensively. Combined with data on the designers' CAD tools, and the increased focus on measuring activity in the design workspace, this proposed tool could also provide a better understanding prototyping and design rationale (Chung & Bañares-Alcántara, 1997) in the whole design process.

# 10 Acknowledgment

This research is supported by the Research Council of Norway through its user-driven research (BIA) funding scheme, project number 236739.

# 11 References

Abras, C., Maloney-Krichmar, D., & Preece, J. (2004). User-centered design. *Bainbridge, W. Encyclopedia of Human-Computer Interaction. Thousand Oaks: Sage Publications*, *37*(4), 445–456.

Ahmed, S., & Christensen, B. T. (2009). An In Situ Study of Analogical Reasoning in Novice and Experienced Design Engineers. *Journal of Mechanical Design*, *131*(11), 111004. https://doi.org/10.1115/1.3184693

Ariff, N. N. A., Badke-Schaub, P., Eris, Ö., & Suib, S. S. S. (2012). *A framework for reaching common understanding during sketching in design teams*.

Auflem, M., Erichsen, J. F., & Steinert, M. (2019). Exemplifying Prototype-Driven Development through Concepts for Medical Training Simulators. *Procedia CIRP*.

Ball, L. J., & Christensen, B. T. (2009). Analogical reasoning and mental simulation in design: two strategies linked to uncertainty resolution. *Design Studies*, *30*(2), 169–186. https://doi.org/10.1016/j.destud.2008.12.005

ball, L. J., & Christensen, B. T. (2018). Designing in the wild. *Design Studies*, *57*, 1–8. https://doi.org/10.1016/j.destud.2018.05.001

Cash, P., Hicks, B., Culley, S., & Salustri, F. (2011). Designer behaviour and activity: An industrial observation method. *DS 68-2: Proceedings of the 18th International Conference on Engineering Design (ICED 11), Impacting Society through Engineering Design, Vol. 2: Design Theory and Research Methodology, Lyngby/Copenhagen, Denmark, 15.-19.08. 2011*, 151–162.

Cash, P., & Maier, A. (2016). Prototyping with your hands: the many roles of gesture in the communication of design concepts. *Journal of Engineering Design*, *27*(1–3), 118–145.

Chakrabarti, A., Morgenstern, S., & Knaab, H. (2004). Identification and application of requirements and their impact on the design process: a protocol study. *Research in Engineering Design*, *15*(1), 22–39. https://doi.org/10.1007/s00163-003-0033-5

Christensen, B. T., & Schunn, C. D. (2007). The relationship of analogical distance to analogical function and preinventive structure: The case of engineering design. *Memory & Cognition*, *35*(1), 29–38.

Chung, P. W. H., & Bañares-Alcántara, R. (1997). Capturing and using design rationale. *AI EDAM*, *11*(2), 89–91. https://doi.org/10.1017/S0890060400001888

Dong, A. (2005). The latent semantic approach to studying design team communication. *Design Studies*, *26*(5), 445–461. https://doi.org/10.1016/j.destud.2004.10.003

Dong, A., Hill, A. W., & Agogino, A. M. (2004). A Document Analysis Method for Characterizing Design Team Performance. *Journal of Mechanical Design*, *126*(3), 378. https://doi.org/10.1115/1.1711818

Dorst, K., & Cross, N. (2001). Creativity in the design process: co-evolution of problem–solution. *Design Studies*, *22*(5), 425–437. https://doi.org/10.1016/S0142-694X(01)00009-6

Erichsen, J. F., Wulvik, A., Steinert, M., & Welo, T. (2019). Efforts on Capturing Prototyping and Design Activity in Engineering Design Research. *Procedia CIRP*, *84*, 566–571. https://doi.org/10.1016/j.procir.2019.04.303

Eris, Ö., Martelaro, N., & Badke-Schaub, P. (2014). A comparative analysis of multimodal communication during design sketching in co-located and distributed environments. *Design Studies*, *35*(6), 559–592. https://doi.org/10.1016/j.destud.2014.04.002

Express - Node.js web application framework. (n.d.). Retrieved 27 March 2019, from http://expressjs.com/

Gero, J. S., & Lindemann, U. (2005). *Human Behaviour in Design 05*. Key Centre of Design Comp & Cogntn.

Goldschmidt, G. (2016). Linkographic Evidence for Concurrent Divergent and Convergent Thinking in Creative Design. *Creativity Research Journal*, *28*(2), 115–122. https://doi.org/10.1080/10400419.2016.1162497

Gonçalves, M., Cardoso, C., & Badke-Schaub, P. (2012). How far is too far? Using different abstraction levels in textual and visual stimuli. *DS 70: Proceedings of DESIGN 2012, the 12th International Design Conference, Dubrovnik, Croatia*.

Hofstede, G. (1984). Cultural dimensions in management and planning. *Asia Pacific Journal of Management*, *1*(2), 81–99. https://doi.org/10/ftmr57

Ishino, Y., & Jin, Y. (2002). Acquiring engineering knowledge from design processes. *AI EDAM*, *16*(2), 73–91. https://doi.org/10.1017/S0890060402020073

Jensen, L. S., Özkil, A. G., & Mortensen, N. H. (2016). Prototypes in engineering design: Definitions and strategies. *14th International Design ConferenceInternational Design Conference*, 821–830. Design Society.

Jensen, M. B., Balters, S., & Steinert, M. (2015). Measuring Prototypes-A Standardized Quantitative Description Of Prototypes And Their Outcome For Data Collection And Analysis. *DS 80-2 Proceedings of the 20th International Conference on Engineering Design (ICED 15) Vol 2: Design Theory and Research Methodology Design Processes, Milan, Italy, 27-30.07. 15*, 295–308. The Design Society.

Jung, M. F., Martelaro, N., & Hinds, P. J. (2015). Using robots to moderate team conflict: the case of repairing violations. *Proceedings of the Tenth Annual ACM/IEEE International Conference on Human-Robot Interaction*, 229–236. ACM.

Jung, M. F., Sirkin, D., Gür, T. M., & Steinert, M. (2015). Displayed uncertainty improves driving experience and behavior: The case of range anxiety in an electric car. *Proceedings of the 33rd Annual ACM Conference on Human Factors in Computing Systems, 2201–2210. ACM.

Kohtala, S., Erichsen, J. A. B., Sjöman, H., & Steinert, M. (2018). Augmenting Physical Prototype Activities in Early-Stage Product Development. *DS 91: Proceedings of NordDesign 2018, Linköping, Sweden, 14th-17th August 2018*.

Larsson, A., Törlind, P., Mabogunje, A., & Milne, A. (2002). *Distributed design teams: embedded one-on-one conversations in one-to-many*. 604–614. Retrieved from http://urn.kb.se/resolve?urn=urn:nbn:se:ltu:diva-40696

Lauff, C. A., Kotys-Schwartz, D., & Rentschler, M. E. (2018). What is a Prototype? What are the Roles of Prototypes in Companies? *Journal of Mechanical Design*, *140*(6), 061102. https://doi.org/10.1115/1.4039340

Lim, Y.-K., Stolterman, E., & Tenenberg, J. (2008). The Anatomy of Prototypes: Prototypes As Filters, Prototypes As Manifestations of Design Ideas. *ACM Trans. Comput.-Hum. Interact.*, *15*(2), 7:1–7:27. https://doi.org/10.1145/1375761.1375762

Mabogunje, A., Eris, O., Sonalkar, N., Jung, M., & Leifer, L. J. (2009). Spider Webbing: A Paradigm for Engineering Design Conversations During Concept Generation. *About Designing: Analysing Design Meetings, J. McDonnell, and P. Llyod, Eds., Taylor & Francis, London, UK*, 49–65.

McAlpine, H., Cash, P., & Hicks, B. (2017). The role of logbooks as mediators of engineering design work. *Design Studies*, *48*, 1–29. https://doi.org/10.1016/j.destud.2016.10.003

Raffaeli, R., Mengoni, M., & Germani, M. (2013). Improving the link between computer-assisted design and configuration tools for the design of mechanical products. *AI EDAM*, *27*(1), 51–64. https://doi.org/10.1017/S0890060412000388

React – A JavaScript library for building user interfaces. (n.d.). Retrieved 27 March 2019, from https://reactjs.org/index.html

Schein, E. H. (1990). Organizational culture. *American Psychologist*, *45*(2), 109–119. https://doi.org/10.1037/0003-066X.45.2.109

Schein, E. H. (2004). *Organizational culture and leadership* (3rd ed). San Francisco: Jossey-Bass.

Schrage, M. (1993). The Culture(s) of PROTOTYPING. *Design Management Journal (Former Series)*, *4*(1), 55–65. https://doi.org/10.1111/j.1948-7169.1993.tb00128.x

Shah, J. J., Smith, S. M., & Vargas-Hernandez, N. (2003). Metrics for measuring ideation effectiveness. *Design Studies*, *24*(2), 111–134. https://doi.org/10.1016/S0142-694X(02)00034-0

Sjöman, H., Erichsen, J. A. B., Welo, T., & Steinert, M. (2017, June). *Effortless Capture of Design Output A Prerequisite for Building a Design Repository with Quantified Design Output*.

Presented at the IEEE International Conference on Engineering, Technology and Innovation 2017, Madeira.

Sonalkar, N., Jablokow, K., Edelman, J., Mabogunje, A., & Leifer, L. (2017). Design whodunit: The relationship between individual characteristics and interaction behaviors in design concept generation. *ASME 2017 International Design Engineering Technical Conferences and Computers and Information in Engineering Conference*, V007T06A009–V007T06A009. American Society of Mechanical Engineers.

Tang, J. C., & Leifer, L. J. (1991). An observational methodology for studying group design activity. *Research in Engineering Design*, *2*(4), 209–219.

Törlind, P. (2007). A Framework for Data Collection of Collaborative Design Research. *Guidelines for a Decision Support Method Adapted to NPD Processes*, 453–454. Paris, France.

Törlind, P., Sonalkar, N., Bergström, M., Blanco, E., Hicks, B., & McAlpine, H. (2009). Lessons learned and future challenges for design observatory research. *DS 58-2: Proceedings of ICED 09, the 17th International Conference on Engineering Design, Vol. 2, Design Theory and Research Methodology, Palo Alto, CA, USA, 24.-27.08. 2009*.

Viswanathan, V., Atilola, O., Esposito, N., & Linsey, J. (2014). A study on the role of physical models in the mitigation of design fixation. *Journal of Engineering Design*, *25*(1–3), 25–43. https://doi.org/10.1080/09544828.2014.885934

Viswanathan, V. K., & Linsey, J. S. (2012). Physical models and design thinking: a study of functionality, novelty and variety of ideas. *Journal of Mechanical Design*, *134*(9), 091004.

Viswanathan, V., Tomko, M., & Linsey, J. (2016). A study on the effects of example familiarity and modality on design fixation. *AI EDAM*, *30*(2), 171–184. https://doi.org/10.1017/S0890060416000056

Wall, M. B., Ulrich, K. T., & Flowers, W. C. (1992). Evaluating prototyping technologies for product design. *Research in Engineering Design*, *3*(3), 163–177. https://doi.org/10.1007/BF01580518

Yang, M. C. (2009). Observations on concept generation and sketching in engineering design. *Research in Engineering Design*, *20*(1), 1–11. https://doi.org/10.1007/s00163-008-0055-0

# Appendix A

Table 2 Table of the 82 prototypes presented in Section 5, sorted chronologically from earliest to latest capture.

| | | | |
|---|---|---|---|
| 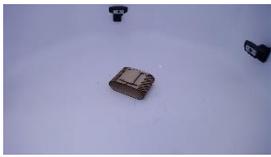 Prototype 1 | 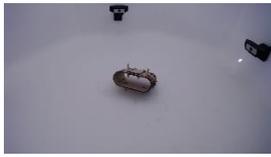 Prototype 2 | 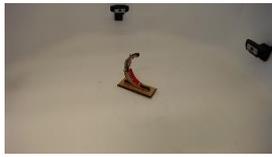 Prototype 3 | 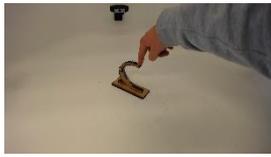 Prototype 4 |
| 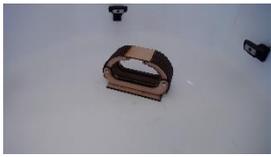 Prototype 5 | 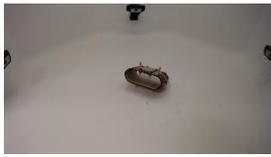 Prototype 6 | 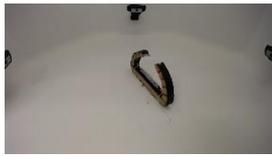 Prototype 7 | 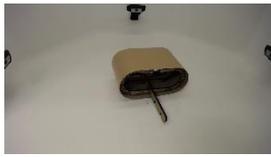 Prototype 8 |
| 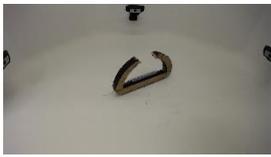 Prototype 9 | 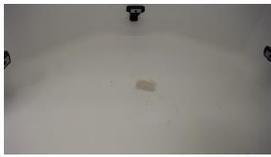 Prototype 10 | 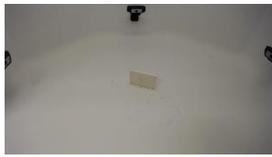 Prototype 11 | 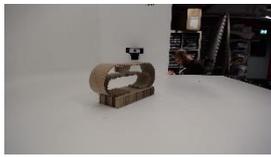 Prototype 12 |
| 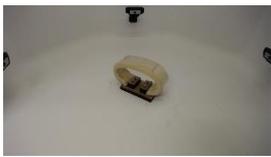 Prototype 13 | 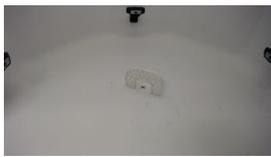 Prototype 14 | 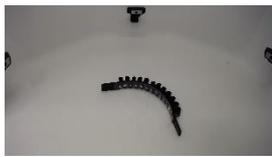 Prototype 15 | 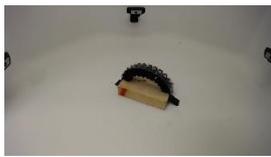 Prototype 16 |
| 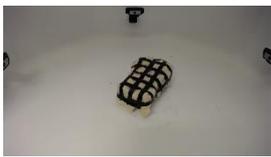 Prototype 17 | 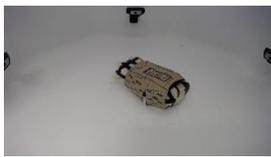 Prototype 18 | 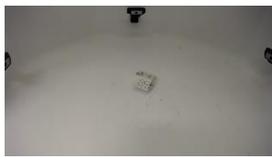 Prototype 19 | 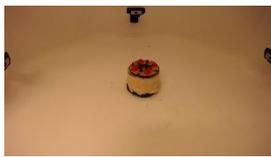 Prototype 20 |
| 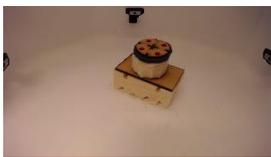 Prototype 21 | 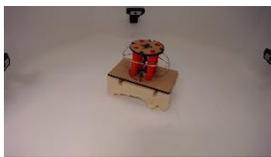 Prototype 22 | 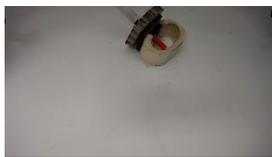 Prototype 23 | 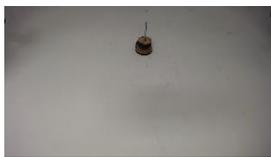 Prototype 24 |
| 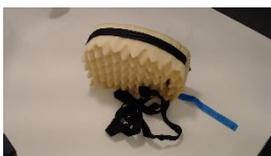 Prototype 25 | 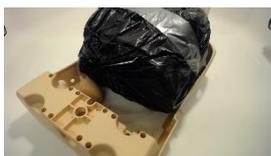 Prototype 26 | 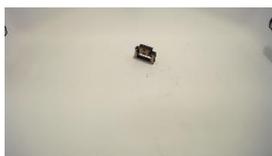 Prototype 27 | 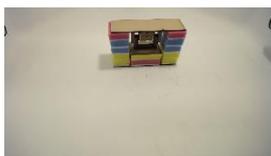 Prototype 28 |

| 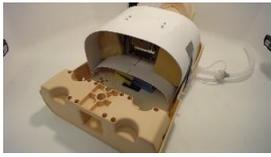 | 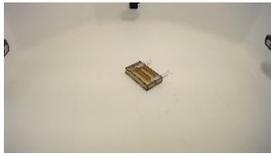 | 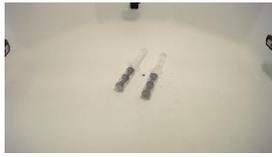 | 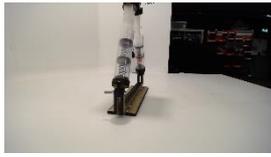 |
|---|---|---|---|
| Prototype 29 | Prototype 30 | Prototype 31 | Prototype 32 |
| 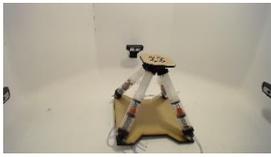 | 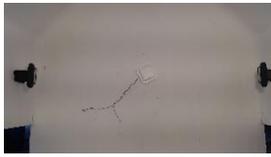 | 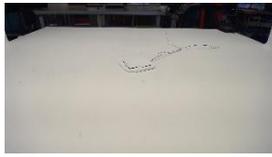 | 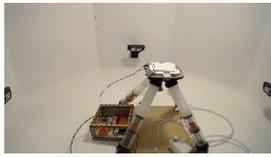 |
| Prototype 33 | Prototype 34 | Prototype 35 | Prototype 36 |
| 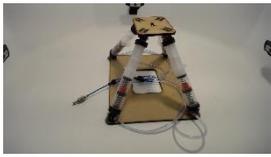 | 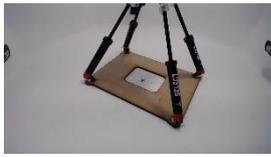 | 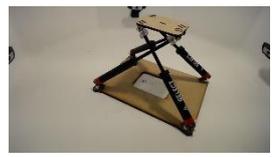 | 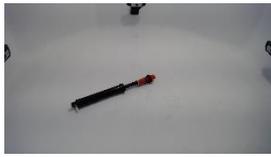 |
| Prototype 37 | Prototype 38 | Prototype 39 | Prototype 40 |
| 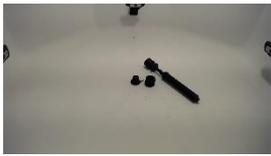 | 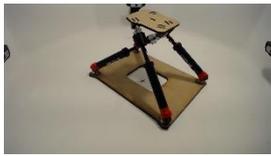 | 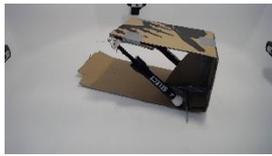 | 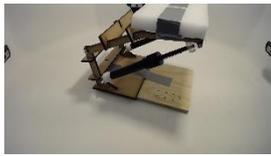 |
| Prototype 41 | Prototype 42 | Prototype 43 | Prototype 44 |
| 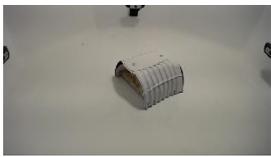 | 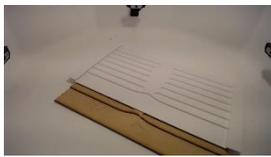 | 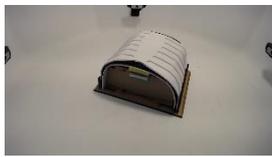 | 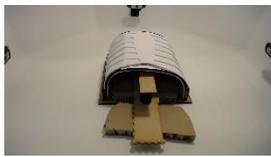 |
| Prototype 45 | Prototype 46 | Prototype 47 | Prototype 48 |
| 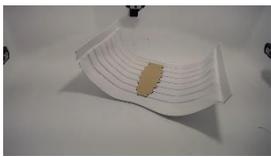 | 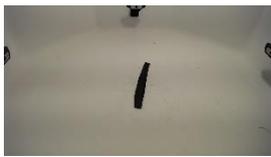 | 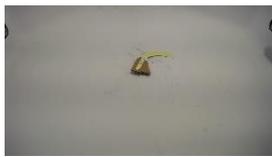 | 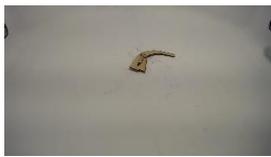 |
| Prototype 49 | Prototype 50 | Prototype 51 | Prototype 52 |
| 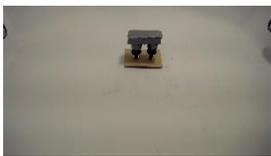 | 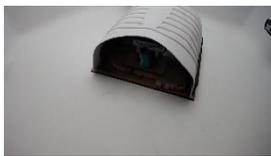 | 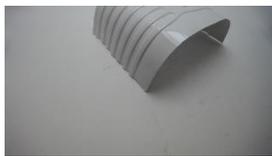 | 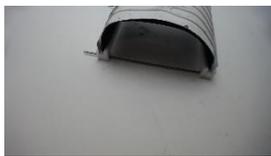 |
| Prototype 53 | Prototype 54 | Prototype 55 | Prototype 56 |
| 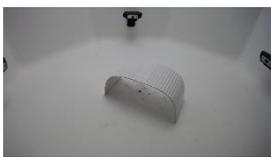 | 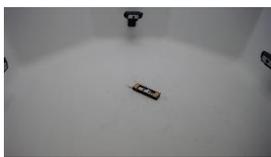 | 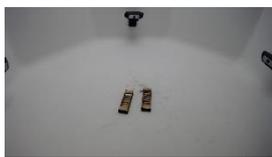 | 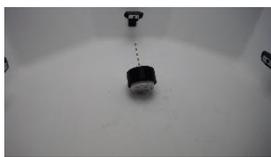 |
| Prototype 57 | Prototype 58 | Prototype 59 | Prototype 60 |

| 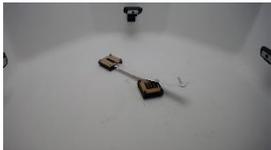 | 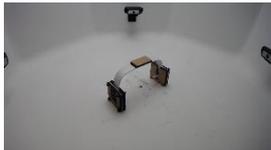 | 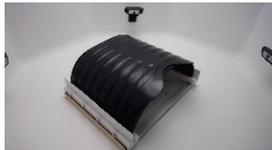 | 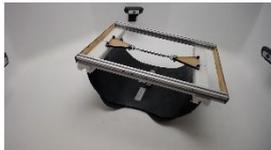 |
|---|---|---|---|
| Prototype 61 | Prototype 62 | Prototype 63 | Prototype 64 |
| 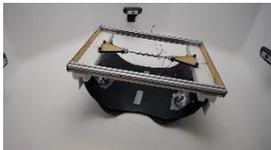 | 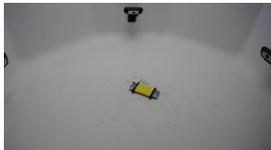 | 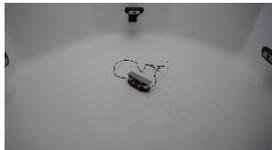 | 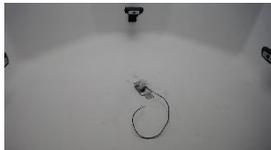 |
| Prototype 65 | Prototype 66 | Prototype 67 | Prototype 68 |
| 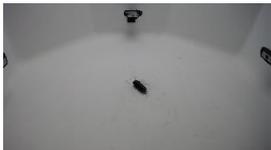 | 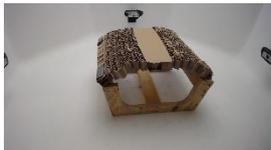 | 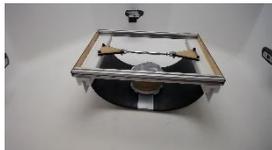 | 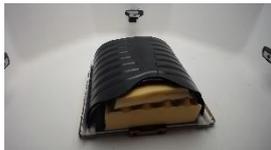 |
| Prototype 69 | Prototype 70 | Prototype 71 | Prototype 72 |
| 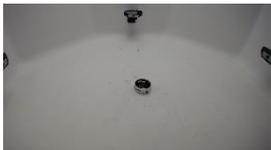 | 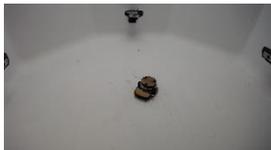 | 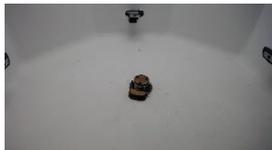 | 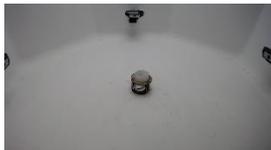 |
| Prototype 73 | Prototype 74 | Prototype 75 | Prototype 76 |
| 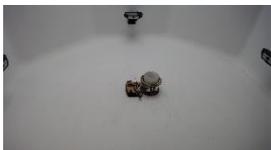 | 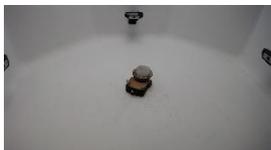 | 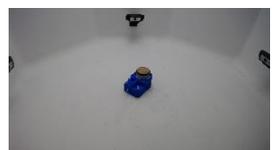 | 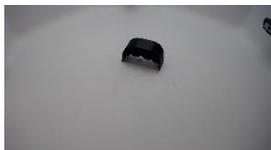 |
| Prototype 77 | Prototype 78 | Prototype 79 | Prototype 80 |
| 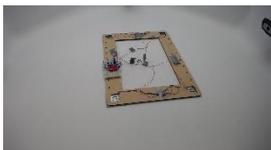 | 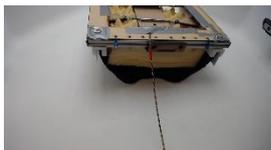 | | |
| Prototype 81 | Prototype 82 | | |